\documentclass[aps,prl,twocolumn,superscriptaddress,reprint,graphicx]{revtex4-1}

\usepackage[utf8]{inputenc}
\usepackage[T1]{fontenc}
\usepackage{lmodern}
\usepackage[english]{babel}
\usepackage{graphicx,pict2e}
\usepackage{hyperref,amsmath}
\usepackage{epstopdf}
\usepackage{color}
\usepackage{multirow}
\usepackage{here}
\usepackage{booktabs}
\usepackage[range-units = single,range-phrase=-,exponent-product=\cdot]{siunitx}
\usepackage{enumitem}

\begin{document}

\author{M. Naumann}
\affiliation{Max Planck Institute for Chemical Physics of Solids, 01187 Dresden, Germany}
\affiliation{Physik-Department, Technische Universität München, 85748 Garchingen, Germany}
\author{F. Arnold}
\author{M. D. Bachmann}
\author{K. A. Modic}
\affiliation{Max Planck Institute for Chemical Physics of Solids, 01187 Dresden, Germany}
\author{P. J. W. Moll}
\affiliation{Max Planck Institute for Chemical Physics of Solids, 01187 Dresden, Germany}
\author{V. Süß}
\author{M. Schmidt}
\affiliation{Max Planck Institute for Chemical Physics of Solids, 01187 Dresden, Germany}
\author{E. Hassinger}
\affiliation{Max Planck Institute for Chemical Physics of Solids, 01187 Dresden, Germany}
\affiliation{Physik-Department, Technische Universität München, 85748 Garchingen, Germany}
\email[Correspondence should be addressed to:]{elena.hassinger@cpfs.mpg.de}

\title{Orbital effect and weak localisation in the longitudinal magnetoresistance of Weyl semimetals NbP, NbAs, TaP and TaAs}

\date{today}

\begin{abstract}

Weyl semimetals such as the TaAs family (TaAs, TaP, NbAs, NbP) host quasiparticle excitations resembling the long sought after Weyl fermions at special band-crossing points in the band structure denoted as Weyl nodes. They are predicted to exhibit a negative longitudinal magnetoresistance (LMR) due to the chiral anomaly if the Fermi energy is sufficiently close to the Weyl points. 
However, current jetting effects, i.e. current inhomogeneities caused by a strong, field-induced conductivity anisotropy in semimetals, have a similar experimental signature and therefore have hindered a determination of the intrinsic LMR in the TaAs family so far.
This work investigates the longitudinal magnetoresistance of all four members of this family along the crystallographic $a$ and $c$ direction. 
Our samples are of similar quality as those previously studied in the literature and have a similar chemical potential as indicated by matching quantum oscillation (QO) frequencies. Care was taken to ensure homogeneous currents in all measurements. 
As opposed to previous studies where this was not done, we find a positive LMR that saturates in fields above 4 T in TaP, NbP and NbAs for $B||c$. Using Fermi-surface geometries from band structure calculations that had been confirmed by experiment, we show that this is the behaviour expected from a classical purely orbital effect, independent on the distance of the Weyl node to the Fermi energy. The TaAs family of compounds is the first to show such a simple LMR without apparent influences of scattering anisotropy.
In configurations where the orbital effect is small, i.e. for $B||a$ in NbAs and NbP, we find a non-monotonous LMR including regions of negative LMR.
We discuss a weak antilocalisation scenario as an alternative interpretation than the chiral anomaly for these results, since it can fully account for the overall field dependence.

\end{abstract}

\maketitle
\section{Introduction}
Topological Weyl semimetals host three-dimensional linear band crossing points of spin-polarised bands in their band structure, so-called Weyl nodes. At the nodes, the quasiparticle excitations behave like relativistic fermions, namely, chiral Weyl fermions. Hence, these condensed-matter systems allow the study of effects which are otherwise only accessible in high-energy physics. In particular, our aim is to investigate the Adler-Bell-Jackiw anomaly or chiral anomaly, an imbalance of the number of left-handed and right-handed Weyl fermions in parallel electric and magnetic fields \cite{Adler69,Bell69} in Weyl semimetals. There, the chiral anomaly contribution to the electrical conductivity should lead to a negative longitudinal magnetoresistance (LMR) \cite{Nielsen83,Son13}. 
Theoretically, its size decreases with increasing energy difference between the Fermi level and the Weyl node and with inter-node scattering, both present in real materials \cite{Son13,Goswami15,Johansson19}.


The prime Weyl semimetal candidate materials where the chiral anomaly might occur is the TaAs family, including TaAs, TaP, NbAs and NbP. These compounds have similar band structures that harbour Weyl nodes, situated within \SIrange{10}{50}{\meV} near the Fermi energy. 
The calculated band structures have been confirmed by angle resolved photoemission spectroscopy (ARPES) and quantum oscillations \cite{Liu15, Xu15,Klotz16, Arnold16TaP, Arnold16TaAs, Arnold19}.

Initial measurements of the LMR indeed revealed it to be negative, which was interpreted as evidence for the chiral anomaly (\cite{Huang15PRX,Zhang16,Li17} and \cite{dosReis16} and references therein).

However, there are two reasons for a reinvestigation of the LMR in this family of compounds. First, it has been shown that applied magnetic fields induce a sizeable conductivity anisotropy in these materials. This results in strong current inhomogeneities when field and current are aligned, so-called current jetting, which disguises the true, intrinsic, LMR \cite{Yoshida76,Pippard89,dosReis16,Arnold16TaP}. In none of the above experiments, current jetting was ruled out. Consequently, the first aim of this work was to detect the intrinsic $\rho_\mathrm{zz}$ in crystals of all members of this family by confirming current homogeneity. 

To achieve this, we follow two routes to a reliable resistance measurement: Injecting the current in a homogeneous manner or increasing the aspect ratio by microstructuring techniques. The latter has been used before \cite{Niemann17} and tiny indications of the chiral anomaly were detected on crystals where the chemical potential was near the Weyl node due to gallium implantation. In contrast, microstructured TaAs for $B||c$ shows a positive and saturating LMR along the crystallographic $c$ direction \cite{Ramshaw18}. 

The second aim of this study was to interpret the results without solely looking for the chiral anomaly. There is a variety of other effects that can explain a negative LMR such as weak localisations \cite{Kawabata80,Altshuler85,Baxter89}, magnetic impurities \cite{Kondo64,Hewson93}, a finite Berry curvature \cite{Goswami15,Dai17,Andreev18}, higher order corrections to the Boltzmann transport equations \cite{Gao17} or a negative Gaussian curvature \cite{Awashima19} of the Fermi surface.

We stress that this study focuses on samples of similar quality and chemical potential to those reported in the literature cited above (see SOM \cite{SOM} tables 1 and 2).

We find the intrinsic LMR to depend strongly on the crystallographic orientation of the current. 
The LMR along the $c$ direction rises and saturates for fields below \SI{4}{\tesla}. The saturation value is in full agreement with the expected orbital MR based on the Fermi surface geometry alone (\cite{Pal10} and SOM chapter 1.6). Along the crystallographic $a$ direction the field dependence is weak and non-monotonic. While in TaP orbital effects can again explain the LMR, in NbP and NbAs the field dependence is in agreement with weak-localisation physics. The longitudinal transport in the TaAs-family can therefore be explained without including any chiral or topological effects, suggesting that these effects are either absent or very small. To our knowledge, a convincing evidence for the chiral anomaly such as the expected increase of chiral currents when the chemical potential is shifted towards the Weyl nodes, is still missing.

\begin{figure}[htp]
 \centering
 \includegraphics[width=1\columnwidth]{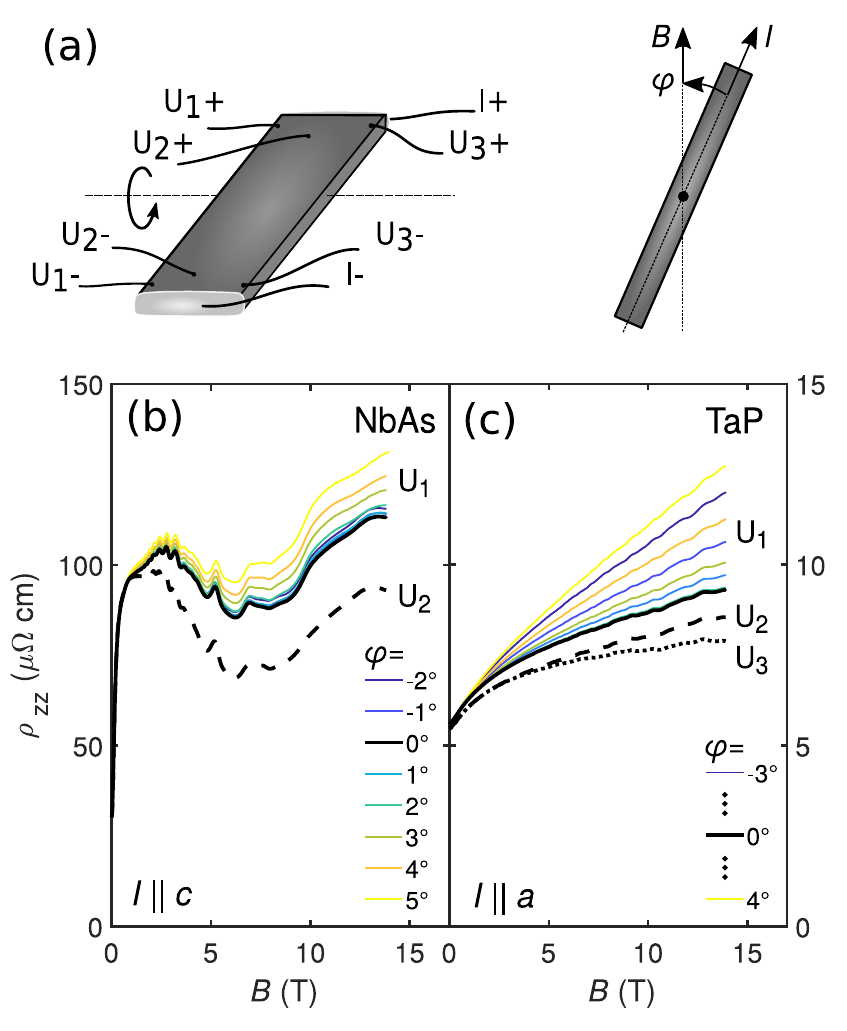}
 \centering
 \caption{Panel \textbf{(a)}: The employed contact geometry for bulk samples and the angle of rotation. \textbf{(b)}: The longitudinal magnetoresistance in NbAs with current along $c$ for some angles around the $B || I$ orientation. \textbf{(c)}: A similar plot for TaP and current along $a$. Different line styles correspond to different contact pairs and the black lines denote the orientation where $B || I$. For better visibility, the angle dependence of only a single contact pair is shown.}
 \label{fig:overview}
 \end{figure}


\section{Experimental Methods}

Crystals of TaAs, TaP, NbAs and NbP were grown by chemical vapour transport using iodine as transport agent. Details of the sample growth are described in the SOM. The crystals showed residual resistivity ratios (RRR = R(\SI{300}{\kelvin})/R(\SI{2}{\kelvin})) of around 5 - 36 depending on sample and the orientation of current with respect to the crystallographic axis. Residual resistivities range from $\SIrange{1.2}{7}{\micro\ohm\centi\metre}$. These residual resistivities and RRRs are in the same range as what was reported in the literature, although they indicate a slightly elevated level of disorder compared to the best samples reported so far (see SOM table 1). 

Most crystals of those materials depart from perfect compensation (equal number of electrons and holes) that is normally expected for semimetals, probably due to crystallographic imperfections below the detection limit of chemical analysis and beyond crystal growth control. Nevertheless, the chemical potential seems to be quite robust since many reports agree on quantum oscillation frequencies in as-grown crystals (see SOM table 2).

TaAs is the only confirmed compound in the family where mirror-symmetric pairs of Fermi surface pockets with non-zero Berry flux through the individual Fermi surfaces are found, an assumed prerequisite for the chiral anomaly \cite{Klotz16, Arnold16TaP, Arnold16TaAs, Arnold19}.

Analysing the Shubnikov-de Haas oscillations in our crystals we find similar levels of the chemical potential as reported in the literature \cite{Shekhar15,Arnold16TaP, Arnold16TaAs,Luo16, Arnold19} (see SOM table 2). In general, QO frequencies $f$ are virtually identical to previous studies in both tantalum compounds ($\Delta f< 1\,\mathrm{T}$) whereas they differ slightly for the Nb compounds ($\Delta f\approx 2\,\mathrm{T}$). 

The crystals were oriented in a Laue diffractometer, cut on a wire or dicing saw and polished into a bar-like shape. The resulting samples had dimensions of around $\SIrange{400}{1000}{\micro\meter}$ in length, $\SIrange{200}{400}{\micro\meter}$ in width and about $\SI{100}{\micro\meter}$ in thickness.

Current jetting primarily arises in materials with a large transverse MR. Compensated semimetals with a high mobility such as Bismuth or the TaAs-family have transverse magnetoresistances of around 1000 in \SI{10}{\tesla} field and are therefore predestined to exhibit this effect.
If the current is injected into the sample through a point-shaped contact, the strong conductivity anisotropy will force it to progress as a jet  along the field direction. This will lead to an apparent negative LMR, since the current is typically diverted away from the voltage contacts in an increasing magnetic field \cite{Yoshida76,dosReis16}. 

For a homogeneous current injection, current contacts were soldered, covering the complete cross section of the samples. 
To get an as local as possible information on the current distribution and homogeneity, up to three pairs of spot-welded voltage contacts distributed over the width of the samples (U1, U2 and U3) were employed. The contact geometry is depicted in figure \ref{fig:overview} (a). 
Previous measurements had shown, that current contacts made from silver paint do not result in a homogeneous current injection \cite{dosReis16}. 
Also, voltage contacts in shape of a stripe of silver paint across the full width of the sample as proposed by \cite{Zhang16} do not detect the intrinsic LMR as a sort of current average (see SOM figure 8 and 9).

Focused ion beam (FIB) microstructured NbAs transport devices were fabricated using a gallium-based Helios NanoLab G3 CX machine. An overview of FIB-based device fabrication of quantum matter can be found elsewhere \cite{MollFIB}. 
It has been shown, that microstructuring of TaAs-type Weyl semimetals using a focused ion beam results in a tantalum or niobium rich surface layer \cite{Bachmann17}. This will become superconducting at temperatures between \SIrange{2}{4}{\kelvin}, which is why the measurements on both NbAs microstructures were performed at \SI{4}{\kelvin}.

Since the conductivity anisotropy and hence current jetting effects are growing in increasing magnetic fields, all voltage contacts typically agree at low fields where they are less dependent on the current contact quality (see the black curves in Fig. \ref{fig:overview} (b) and (c)). At higher fields, deviations between different contacts were visible in most cases. For qualitatively similar field dependencies with only small differences as in Fig. \ref{fig:overview} (b) and (c), the current distribution was deemed sufficiently homogeneous and we will refer to the $\rho_\mathrm{zz}$ as 'intrinsic'.

It should be noted that even a perfectly homogeneous current injection is still susceptible to current jetting effects, if the sample is not perfectly aligned in the external magnetic field. Under these circumstances, the 'jet' will be as large as the sample cross section, passing through the sample in the direction of the misaligned magnetic field \cite{Yoshida76}. In order to spot this, field sweeps were performed at various angles $\varphi$ around the parallel orientation (see figure \ref{fig:overview}). The rotation around $\varphi$ was done in-situ at $\SI{2}{\kelvin}$ in a $\SI{14}{\tesla}$ QD Physical-Property Measurement System (PPMS) using a QD rotator probe.  Magnetic field sweeps were performed from \SI{-14}{\tesla} to \SI{14}{\tesla} (or vice versa) and data was subsequently symmetrised to exclude any Hall contribution to the signal. In these rotation studies, a clear sign of current jetting is when a contact showing a strong positive magnetoresistance in high fields would turn to a \textit{"negative resistance"} upon rotating only a few degrees. The alignment around the second angle $\vartheta$ was done by eye under a microscope, leading to uncertainties of a few degrees.

For both $B||a$ and $B||c$, the symmetry of the system results in a conductivity tensor of reduced form with $\sigma_{i\mathrm{z}} = \sigma_{\mathrm{z}i} = 0$ with $i=\mathrm{x,y}$, where $\sigma_\mathrm{zz}$ can be easily retrieved by inverting $\rho_\mathrm{zz}$. In contrast, if the system is tilted slightly off a main crystallographic axis, the symmetries preserving this form are broken and $\sigma_\mathrm{zz}$ can only be extracted by a full tensor inversion of $\hat{\rho}$ and vice versa. This leads mainly to an admixture of the transverse magnetoconductivity to the signal that smoothly grows with the angle. Both niobium compounds with current along $a$ had an increasing resistance in high fields with this angle dependence. In that case, only the low-field behavior, where the transverse MR goes to zero, was ascribed to the LMR.

For reasons of clarity the main text contains mainly results from one voltage contact pair on each sample. The reader is referred to the SOM, where the full results are shown.

\section{Possible phenomena causing a magnetoresistance}

In order to aid the interpretation of the data shown hereafter, we would like to give an overview over effects contributing to an LMR.
\begin{enumerate}[label=(\roman*)]
  \item The orbital LMR of a free electron gas with spherical Fermi surface is known to be zero \cite{Pippard89}. However, a particular corrugation of the Fermi-surface gives rise to a finite LMR caused by orbital effects \cite{Yoshida76,Pal10}. The model detailed in \cite{Pal10} is based on a Boltzmann approach to the transport problem and only takes the geometry of the Fermi surface into account. The scattering time is isotropic. In a nutshell, if the average Fermi velocity component in the direction of field and current varies along a cyclotron orbit, there will be a finite LMR and the expected change in conductivity in the high field limit $\delta \sigma/\sigma_0 = (\sigma(B=\infty)-\sigma(0))/\sigma(0)$ can be obtained. Here we calculate $\delta \sigma/\sigma_0$ of all compounds and in both crystallographic current/field directions using the Fermi surfaces from DFT results that showed good agreement with quantum oscillations \cite{Arnold16TaP, Arnold16TaAs, Klotz16,Arnold19} (see the supplemental material). Results are presented in table 3 in the SOM. How fast the saturation value is reached depends on the mobility $\mu$ and the orbital magnetoresistance is given by 
  \begin{align}
    \sigma_\mathrm{orb}(B) &= \frac{ne\mu}{1+(\mu B)^2}, 
  \end{align}
  with the charge carrier density $n$. Since this is a decreasing function in $B$, the magnetoresistance is positive. 
  
  Despite this, there is a recent theoretical prediction, that a negative Gaussian curvature of the Fermi surface can cause a negative LMR \cite{Awashima19}.
   
  \item The chiral anomaly contributes a positive term to the total longitudinal conductivity when magnetic and electric fields are parallel \cite{Nielsen83,Son13}. In the semiclassical regime
  \begin{align}
    \sigma_\mathrm{CA} &= C \cdot B^2
  \end{align}
  with $C$ being a constant in field \cite{Son13, Goswami15}. 
  
In the simplest possible model, the size of this contribution $C$ decreases with the distance of the Fermi energy to the Weyl nodes by $|E_\mathrm{Weyl}-E_\mathrm{F}|^{-2}$ \cite{Son13}, but the exponent decreases to at least -3 in a more realistic model \cite{Johansson19}. In this model, a distance of less than \SI{5}{\meV} is sufficient to make the chiral conductivity contribution negligible. 
Note that calculations for the real band-structures are missing, implying absolute numbers to be different in real materials.

\item  Electron-self interference effects, also called weak (anti-) localisation can induce both positive and negative MR. These effects occur, when the electron dephasing time $\tau_\phi$ is larger than the scattering time $\tau$. Whether a weak localisation or a weak antilocalisation occurs depends on the symmetry of the system. In the materials investigated here, the existence of spin-polarised Fermi surfaces breaks the spin rotational symmetry and therefore only a weak antilocalisation (WAL) is applicable. The additional conductivity contribution is different for Weyl-systems, $\sigma_\mathrm{WWAL}(B)$ \cite{Lu15}, and those with a 'trivial', non-chiral electronic structure, $\delta \sigma_{WAL} (B)$ \cite{Kawabata80,Altshuler85,Baxter89}. The magnetic field dependence of the conductivity of a Weyl weak antilocalisation (WWAL) can be expressed as

\begin{align}
  \delta\sigma_\mathrm{WWAL}(B) &= \sigma_\mathrm{WWAL}(B) - \sigma_\mathrm{WWAL}(0)\nonumber \\
  \sigma_\mathrm{WWAL}(B) &= \frac{2e^2}{(2\pi)^2 h}  \int_0^{1/l} \mathrm{d} x \left [ \psi_{l_B}(l) - \psi_{l_B}(l_\phi) \right ]\\
  \psi_{l_B}(z) &= \Psi\left ( \frac{l_\mathrm{B}^2}{z^2} + l_\mathrm{B}^2x^2 + \frac{1}{2}\right )\nonumber\\ 
\end{align}

Here, $l$ denotes the mean free path, $l_\phi$ the dephasing length, $l_\mathrm{B}=\sqrt{\hbar/(4eB)}$ the magnetic length and $\Psi$ the Digamma function \cite{Lu15}. 

The correction to the conductivity due to a 'trivial' weak antilocalisation in 3D is given by 
\begin{align}
  \sigma(B) &= \sigma(0) + \delta\sigma_{WAL}(B)\nonumber\\
  \delta \sigma_{WAL} (B) &= -\frac{e^2}{2\pi^2\hbar} \sqrt{\frac{eB}{\hbar}} \times...\nonumber\\
    ... &  \left [ \frac{1}{2} f_3 \left ( \frac{B}{B_\phi} \right ) - \frac{3}{2} f_3 \left ( \frac{B}{B_2} \right ) \right ]\label{eq:3DWAL}\\
  B_2 &= B_\phi + \frac{4}{3}B_{so}\nonumber\\
  f_3 (x) &= \sum_{n=0}^\infty \left [ 2\left (n+1+1/x \right )^{1/2}  ... \right .\nonumber\\
   ...  & \left . -2\left (n+1/x \right )^{1/2} -\left (n+1/2+1/x \right )^{-1/2} \right ]\nonumber
\end{align}

The quantities $B_{\phi,so}$ are parameters which are determined by the dephasing or spin-orbit scattering times via the relation $B_i = \hbar /(4eD\tau_i)$ with $i=(\phi,\mathrm{so})$ and $D$ being the diffusion constant. Apparently, the WWAL will result in a negative conductivity contribution, ie. a positive magnetoresistance. In contrast, the 'trivial' 3D WAL will result in a negative magnetoresistance at high fields which may be positive at low fields depending on the particular choice of the spin-orbit scattering time $\tau_\mathrm{so}$ relative to the dephasing time $\tau_\phi$ \cite{Kawabata80,Altshuler85,Baxter89}.

It should be noted, that these models are the full results of the theory. WAL type of effects have already been discussed in conjunction with a chiral anomaly in both arsenides \cite{Huang15PRX,Zhang16,Li17} based on approximations such as e. g. $\rho \propto \sqrt{B}$, which is only true for a rather large separation of $l$ and $l_\phi$.

\item Magnetic impurities 
can lead to a negative magnetoresistance, known from e.g. Kondo systems: Upon alignment of the magnetic moments by the external magnetic field, the disorder is decreased and so is the resistance \cite{Hewson93}. We found an upper threshold of \SI{100}{ppm} magnetic impurities in our samples based on magnetization results  (see section 1.5 in the SOM). Traditional systems with similar impurity concentrations, such as Au\textit{Fe}, host a sizeable Kondo effect \cite{Kondo64}. However, the bandwidth and density of states at the Fermi energy in semimetals is appreciably lower. Since the Kondo temperature depends linearly on the former and exponentially on the latter, it should be suppressed by several orders of magnitude. Therefore we rule out that magnetic impurities influence the LMR in this study.

\item Other reasons for a longitudinal MR are anisotropic scattering, macroscopic inhomogeneities, barrier inhomogeneities in superlattices or a modification of the density of states in field (see \cite{Pal10} and the references therein) but will not be discussed here. The same applies to second order band structure effects including the Berry curvature \cite{Andreev18}. As shown below, the effects that are enumerated above are sufficient to describe the experimental results.

\end{enumerate}

From this list, it can be seen that a negative LMR may be caused by a chiral anomaly, a weak antilocalisation, magnetic impurities or band structure effects such as a negative gaussian curvature. Therefore, the implication used occasionally that a negative LMR proves a material to be topologically non-trivial is wrong.

\section{$c$ - axis transport}

\begin{figure}[htb]
\centering
\includegraphics[width=0.93\columnwidth]{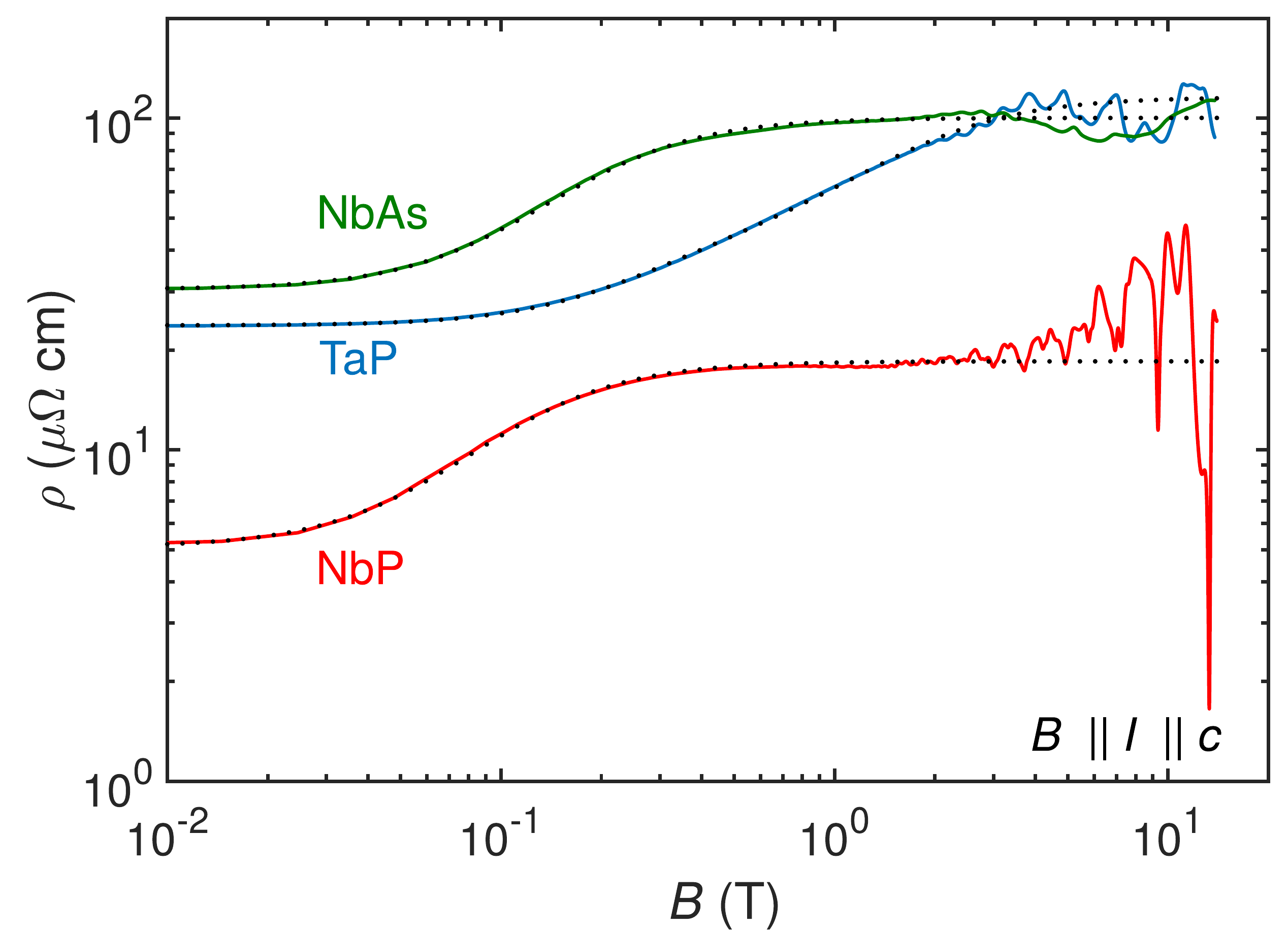}
\centering
\caption{The longitudinal magnetoresistance for TaP, NbAs and NbP with current along $c$. The fits to the orbital model are shown as black dotted line.}
\label{fig:Iparac}
\end{figure}

Figure \ref{fig:Iparac} shows the longitudinal magnetoresistance of TaP, NbP and NbAs for current along the crystallographic $c$ direction. In all three compounds, the resistivity rises and saturates at fields between \SIrange{2}{4}{\tesla}. In TaP, higher fields are necessary to reach the saturation than in the niobium compounds. All signals are superimposed by quantum oscillations; in NbP they are extremely pronounced and their size is of the same order as the rise in resistivity in field. In this paper, we concentrate on the background MR. Since all contact pairs agree qualitatively in the measurements shown here, this is the intrinsic LMR (see SOM, figures 1, 6, 10). This is also supported by the rather small angular dependence of the initial increase.

The TaAs crystals at our disposal were too small along the $c$ axis to make samples suitable for meaningful measurements.


There are two possible effects contributing to a positive LMR: A WAL and orbital effects. Since the inherent negative LMR of the WAL is clearly absent, only orbital effects can be at the origin of the increasing resistivity.

The fit presented as dotted line in figure \ref{fig:Iparac} takes two orbital contributions (from warped parts of the Fermi surface of the electron and the hole pocket) and a field independent conductivity channel $\sigma_0$ (from unwarped parts of the FS) into account. 
\begin{align}
  \sigma (B) &= \sigma_\infty + \sigma_\mathrm{orb1}(B, \mu_1, n_1) + \sigma_\mathrm{orb2}(B, \mu_2, n_2).\label{eq:Iparc_model}\\
    \rho (B) &= \sigma(B)^{-1}
\end{align}
In this model, $\mu_{1/2}$ and $n_{1/2}$ are the electron / hole mobility and carrier densities and were treated as free parameters. In both Nb compounds, only one orbital term is needed, which is why $n_2$ was set equal zero: In NbAs one band dominates the $c$-axis transport (see SOM table 3) and in NbP both contributing bands have similar mobilities, making them indistinguishable \cite{Moritz}. In the Nb compounds, the fits were performed between \SIrange{0}{4}{\tesla}, in TaP between \SIrange{0}{6}{\tesla}, since the resistivity saturates at higher fields. The cutoff was introduced to avoid any influence of the pronounced quantum oscillations and to keep more weight on the low-field increase, which was also the reason why it was performed on log-scaled data. 
The resulting parameters can be found in SOM table 5. The relative change of conductivity in field $\delta\sigma/\sigma_\mathrm{0,exp}$ is well in accordance with the expected one from the DFT Fermi-surface alone, $\delta\sigma/\sigma_\mathrm{0,th}$ \cite{Pal10} (see SOM table 3). 
Also, the mobilities agree well with literature values of the $a$-axis mobilities $\mu^a$, as determined from a zero-field Drude model or two-band Hall fits \cite{Arnold16TaP,Luo16,Klotz16,Arnold16TaAs}. 
The relatively simple model by Pal and Maslov describing semiclassical orbital effects of an irregular Fermi surface with isotropic scattering times seems to capture the essential physics of the LMR in NbAs, NbP for $B||c$ and in both field directions in TaP. Small angle scattering must also be negligible, since it increases the MR \cite{Pippard64}. This result also implies that the chiral anomaly contribution is either absent or very small in contradiction to previously published reports \cite{Li17}. Note that the energy distance of the Weyl node to the Fermi level $|E_\mathrm{Weyl}-E_\mathrm{F}|$ from the DFT band structures is smaller in NbAs ($5\pm 5$\,meV) than in NbP ($10\pm 5$\,meV), TaP (13\,meV) and TaAs (13\,meV) \cite{Arnold16TaP, Arnold16TaAs,Klotz16,Arnold19}. However, there appears to be no correlation between $|E_\mathrm{Weyl}-E_\mathrm{F}|$ and the occurrence or size of a negative LMR.

\section{$a$ - axis transport}
\begin{figure}[htb]
\centering
\includegraphics[width=0.93\columnwidth]{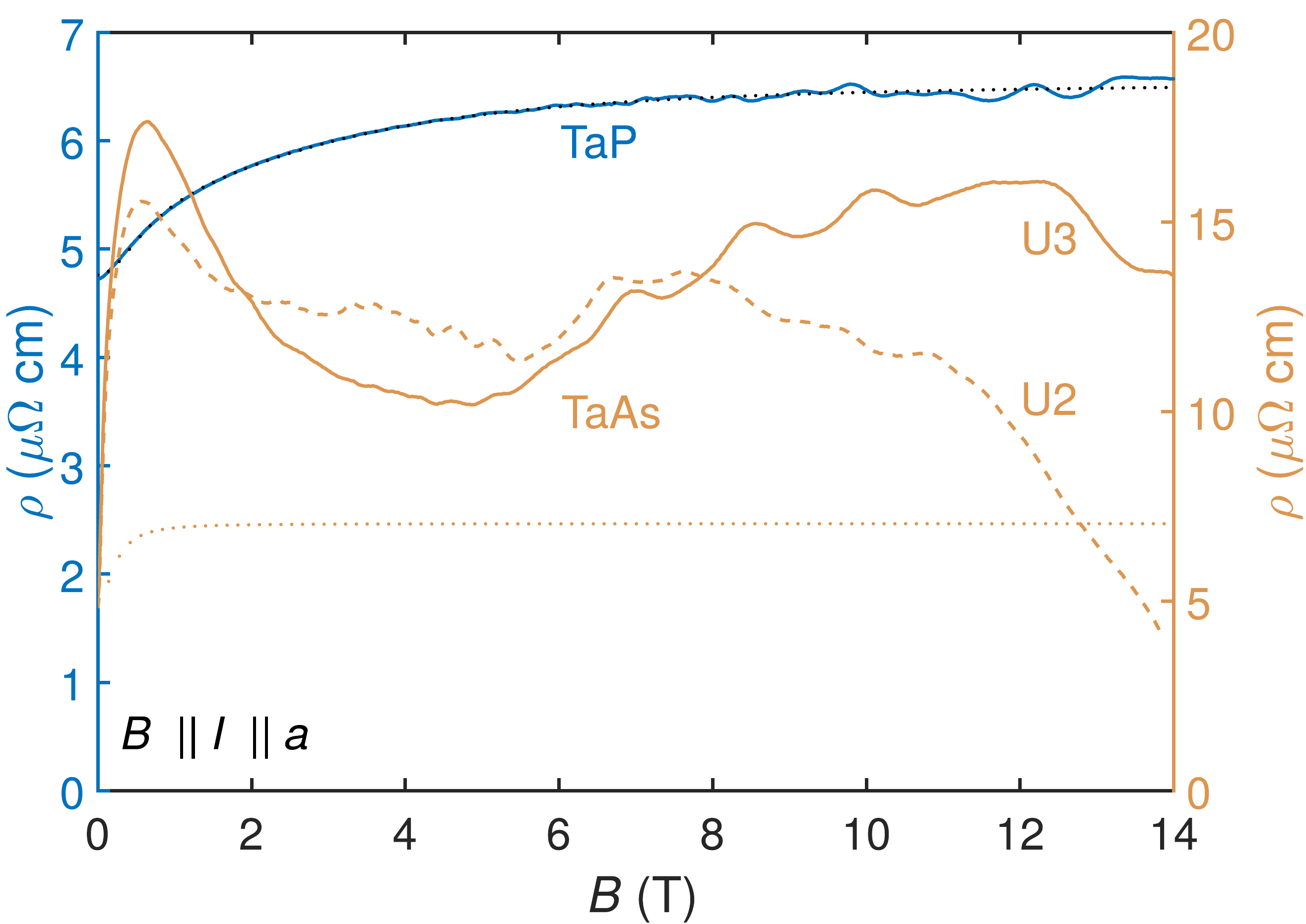}
\centering
\caption{The longitudinal magnetoresistance for both tantalum compounds with current along $a$. The lower orange dotted line indicates the expected orbital LMR in TaAs with a hypothetical \SI{5}{\per\tesla} mobility as an illustration.}
\label{fig:tantal_ipara}
\end{figure}

The $a$-axis transport in \textbf{TaP} (Fig. \ref{fig:tantal_ipara}) strongly resembles the $c$-axis transport. No sign of any negative magnetoresistance is present. Here, all contact pairs agreed rather well (see SOM, figures 2 and 3). Therefore the resistance curve shown reflects the true longitudinal magnetoresistance in TaP. A second sample gave the same qualitative field dependence (SOM figure 2).The expected conductivity change for the $a$-axis transport in TaP (see Fig. \ref{fig:tantal_ipara}) based on the FS geometry is also in agreement with the experimental results. Fit results are presented in the SOM, table 5. Again, the high-field saturation of the magnetoresistance agrees with the expectations based on the Fermi surface corrugation.

As expected for this field orientation, the mobilities are lower: Due to the asymmetry of the Fermi-surface, the effective masses are a factor of about 7 higher, while the mobility anisotropy we find is about three.

For \textbf{TaAs}, both voltage contacts are shown in figure \ref{fig:tantal_ipara}.  Given the different field dependence of the resistivity curves, we deem the current homogeneity insufficient to make reliable statements regarding the intrinsic LMR.  
However, both curves concordantly show a hump at low field followed by a saturation at high fields, likely intrinsic features of the MR. It resembles the results shown in \cite{Ramshaw18}, where $c$-axis transport data is presented.

The initial increase though is beyond what is expected from the orbital contribution, as can be seen by comparing the data and the estimated orbital background (dotted line) in figure \ref{fig:tantal_ipara}. Strong small-angle scattering can enhance the longitudinal MR \cite{Pippard64}. However, the similarity of the hump in TaAs with the one in NbAs hints to weak-antilocalisation physics as discussed in the next paragraph.

\begin{figure}[htb]
\centering
\includegraphics[width=0.93\columnwidth]{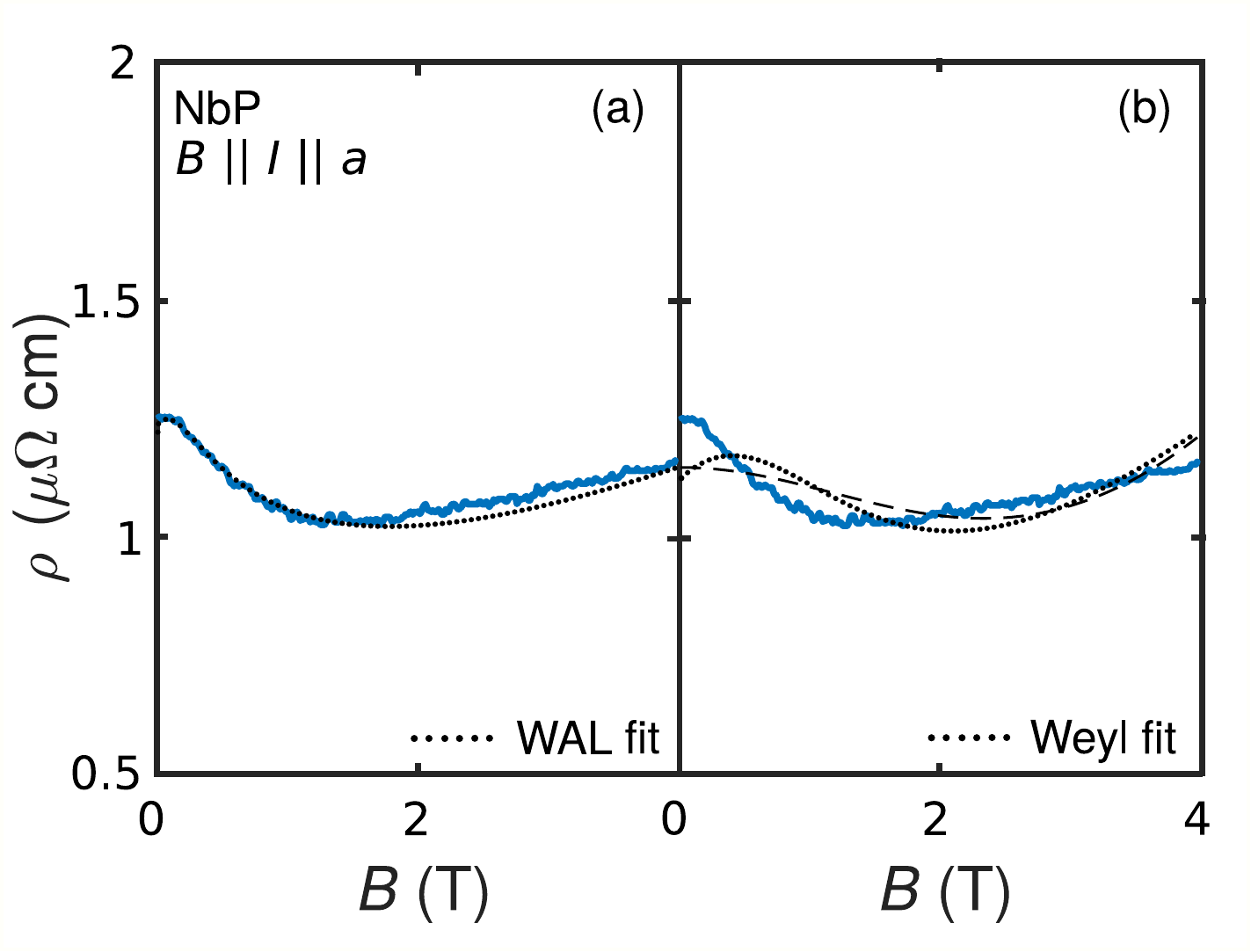}
\centering
\caption{The longitudinal magnetoresistance of NbP with current along $a$. The overall changes in resistivity are rather small. The black dotted and dashed lines are fits according to (a) the WAL model (dotted: linear $f_{BG}$, dashed: quadratic $f_{BG})$, Eq. \ref{eq:niob_fitf}, and (b) the Weyl model, equation \ref{eq:Weyl}, with the Weyl WAL contribution constricted to zero.}
\label{fig:NbP}
\end{figure}

In \textbf{NbP}, figure \ref{fig:NbP}, we can see the magnetoresistance to be rather small, only of the order of \SI{25}{\percent} in total. The signal decreases at low fields and increases almost linearly afterwards. Since the downturn at low fields shows up the same way in all three contact pairs (see SOM, figure 12), we believe this to be of intrinsic origin. In higher fields, signs of current jetting appear, such as disagreement between contact pairs and negative voltages in the angle dependence (see SOM fig. 12 and 13). For the following interpretation the focus is on the low-field behavior whereas the high field increase is attributed to a misalignment of the sample in field.

\begin{figure}[htb]
\centering
\includegraphics[width=0.93\columnwidth]{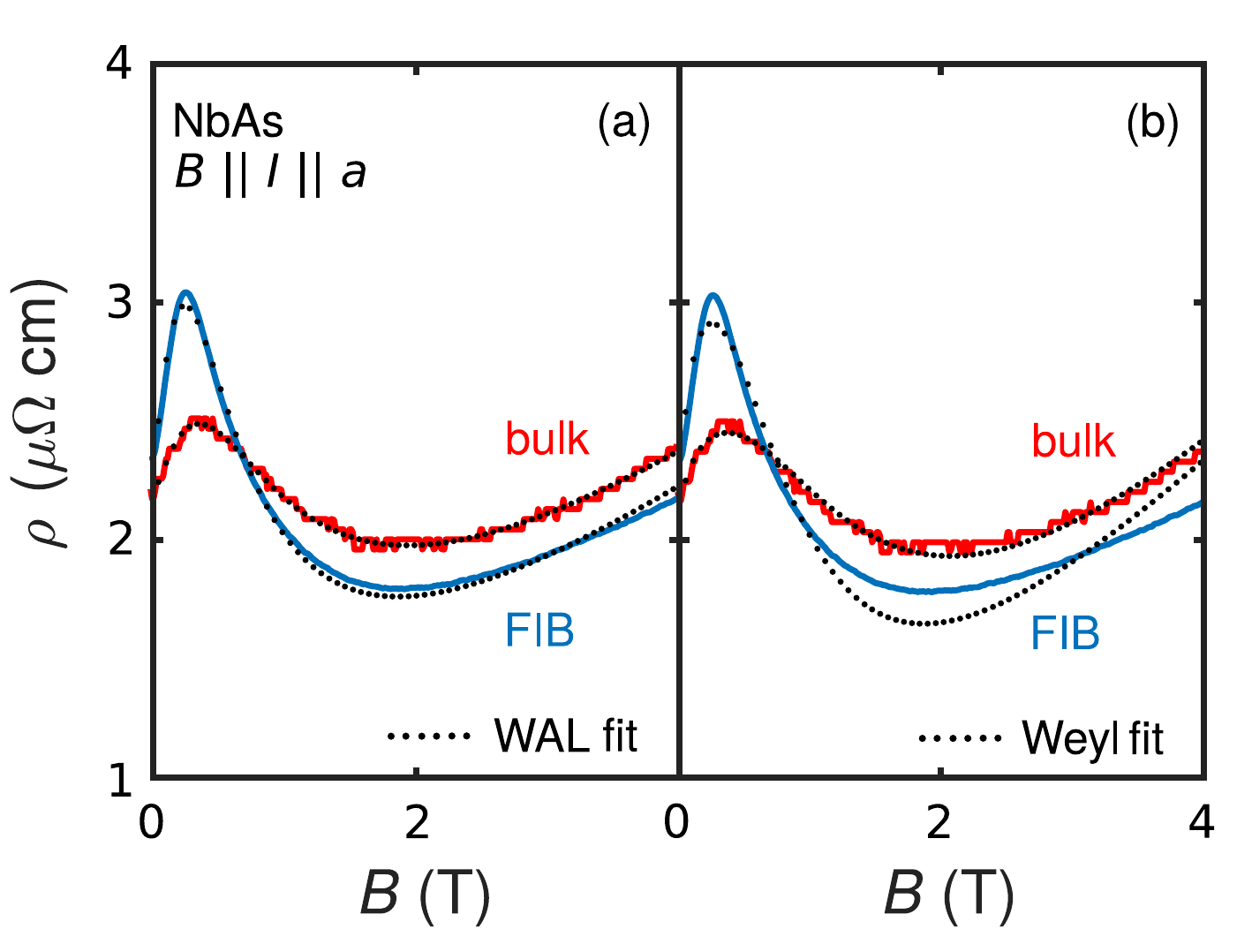}
\centering
\caption{The longitudinal magnetoresistance of both a bulk (red) and a microstructured sample (blue) of NbAs. The black dotted lines denote fits according to (a) to the weak antilocalisation model, equation \ref{eq:niob_fitf}, and (b) according to the 'Weyl-model', Eq. \ref{eq:Weyl}. }
\label{fig:NbAs_two}
\end{figure}

The overall changes in resistivity are slightly higher in \textbf{NbAs} as shown in figure \ref{fig:NbAs_two}. At lowest fields, there is an increase in resistivity while the signal beyond \SI{0.3}{\tesla}  resembles the one of NbP. We also attribute the high-field resistivity increase to a misalignment of the sample's main axis with respect to the magnetic field. Although we only had one working voltage contact on this FIB microstructure, there is a number of reasons why we consider the signal to be reliable at low field: Measurements on a second microstructured sample with $I || c$ present no difference between the different voltage contacts up to about \SI{1}{\tesla} and an almost field independent difference of about \SI{20}{\percent} at fields greater \SI{4}{\tesla} (see figure \ref{fig:overview} B). Current jetting effects are subdominant once the aspect ratio is of the order of the square root of the conductivity anisotropy ratio $\sqrt{\sigma_{||}/\sigma_\perp}$ \cite{Yoshida76}. We estimate this using resistivities to be $\sqrt{\rho_{B||a, I||c}/\rho_{B||I||a}}=7$ at \SI{0.5}{\tesla} and 20 at \SI{4}{\tesla}. Since the microstructure has dimensions of \SI{2.75}{\micro\meter} x \SI{4.83}{\micro\meter} x \SI{88}{\micro\meter} (w x t x l), the samples' aspect ratio of about 18 is close to the conductivity anisotropy ratio.

The fact that the overall $a$-axis LMR in both Nb compounds is smaller than the $c$-axis one is in perfect agreement with the expected orbital MR based on the FS shape, which suggest a tiny $\delta \sigma/\sigma_{\mathrm{0}}= \num{0.075}$ in NbAs and \num{0.097} in NbP (see SOM, table 3). As a consequence, we neglect the orbital effect in the following analysis.
Both materials show a strong field dependence in low fields including regions of negative magnetoresistance. 
In the following, we therefore discuss two possible scenarios for the non-monotonic (NbAs) or negative (NbP) behaviour in low magnetic fields:  
In the "WAL" scenario, a trivial 3D weak antilocalisation $\delta\sigma_{WAL}$ describes the full low-field behavior.
\begin{align}
  \rho_\mathrm{WAL} (B) &= \left [\sigma_0 + A\cdot \delta\sigma_{WAL}(B_\phi, B_\mathrm{so}) \right ]^{-1} ...\nonumber\\ 
  & ... + f_\mathrm{bg}(B)\label{eq:niob_fitf}
\end{align}

In the "Weyl" scenario, Weyl fermions dominate the LMR causing a negative WWAL correction to the conductivity $\delta\sigma_\mathrm{WWAL}$ and at higher fields a positive chiral anomaly contribution $\sigma_\mathrm{CA}$. 
\begin{align}
\rho_\mathrm{Weyl} (B) &= \left [ \sigma_0 + A\cdot\delta\sigma_\mathrm{WWAL}(l,l_\phi) + \sigma_\mathrm{CA} \right ]^{-1}...\nonumber\\
  & ... + f_\mathrm{bg}(B)\label{eq:Weyl}
\end{align}

Both models include a constant conductivity contribution $\sigma_0$ and a function $f_\mathrm{bg}(B)$ accounting for the slowly increasing resistivity background at higher fields assumed to be stemming from a small misalignment. A simple linear function $f_\mathrm{bg}(B) = d\cdot B$ was used as simplest possible function fulfilling the requirement of $f_\mathrm{bg}(B\to 0) \to 0$. Since this caused a hump at low fields in the fit, a quadratic background was also tried in NbP (see dotted (linear) vs. dashed (quadratic) lines in figure \ref{fig:NbP}). In NbAs, this functional difference did not change the fit parameters significantly. 

Results of these fits can be seen as black dotted lines in figure \ref{fig:NbP} and figure \ref{fig:NbAs_two}. The resulting fit parameters are given in full in the supplemental material whereas we discuss only the crucial ones here. 

The fit according to the weak antilocalisation scenario describes the low-field behaviour of both NbP and NbAs very well. We now check if the model is applicable at all. This is done by verifying the hierarchy of timescales  $\tau\ll\tau_\mathrm{so}\ll\tau_\phi$, with the electron dephasing time $\tau_\phi$, spin-orbit scatteriung time $\tau_\mathrm{so}$ and total scattering time $\tau$. From the fit parameters $B_\phi = \SI{9.7}{\milli\tesla}$ (\SI{17}{\milli\tesla}) and $B_\mathrm{so} = \SI{72}{\milli\tesla}$ (or \SI{23}{\milli\tesla}) for NbAs (or NbP) we extract the dephasing and scattering times $\tau_\phi$ and $\tau_\mathrm{so}$ by using an estimation of the diffusion constants $D$ (for details of the extraction see the SOM, 1.9.1). For NbAs we find the inequality $\tau\ll\tau_\mathrm{so}\ll\tau_\phi$ is fulfilled for the electron band such as $\SI{0.5}{\pico\second} < \SI{3}{\pico\second} < \SI{20}{\pico\second} $, whereas in NbP it barely is $\SI{1.0}{\pico\second} < \SI{5}{\pico\second} < \SI{8}{\pico\second} $. It should be kept in mind that scattering times are typically associated with uncertainties up to one order of magnitude.

With this large error bar in mind, we can state that the necessary hierarchy for a WAL to occur ($\tau\ll\tau_\mathrm{so}\ll\tau_\phi$) is mainly fulfilled. The size of the MR we observe, is large compared to other reports of weak localisation MR \cite{Kawabata80,Baxter89}. Since this study is intended to investigate the field dependence of the LMR, a detailed temperature dependence, that would be needed to confirm the WAL scenario, is beyond the scope of this work. An open question is also, why the WAL appears only in the $a$-axis transport and not in the $c$-axis.

In figures \ref{fig:NbP} b and \ref{fig:NbAs_two} b, a fit to the LMR in the Weyl scenario is shown. In NbP, the Weyl fit is performed without the WWAL contribution, since there is no low-field increase observable. This model cannot describe the experimental data, both with a linear and quadratic $f_\mathrm{BG}$, as shown as dotted (linear) and dashed (quadratic) lines in figure \ref{fig:NbP}. The agreement is better in NbAs, although it is clearly worse compared to the WAL model (Fig. \ref{fig:NbAs_two}). The fit parameters (table 6 in the SOM) indicate that the mean free path is of the same size as the dephasing length $l\simeq l_\phi$ and consequently, that the necessary separation of the two length scales is absent.

Our results presented here therefore reveal a number of observations which cannot be reconciled with the interpretation of a chiral anomaly being responsible for the negative LMR observed in these materials for $B||a$:
\begin{itemize}
  \item In NbAs, the Weyl WAL fit of the upturn in resistivity at low fields yields unphysical fit parameters. But other possible origins of the upturn in the Weyl picture are absent since the orbital contribution is too small.  Also, the presence of a trivial WAL responsible for the positive LMR would imply the presence of a negative LMR as well. The presence of two effects causing a negative LMR would reduce the individual contribution of the chiral anomaly to a rather low level and make its distinction and study even more cumbersome.
  \item In NbP no good agreement between the Weyl model and the data could be achieved.
  \item The possible influence of the negative Gaussian curvature of the Fermi surfaces has not been investigated or discussed so far. The 'banana' shaped Fermi surfaces present in all materials of this family have large regions of negative gaussian curvature (positive curvature in one direction and negative in the perpendicular direction) this might very well contribute to the negative LMR \cite{Awashima19}. 
\end{itemize}

Given that the relatively simple picture of a weak antilocalisation is in agreement with the low-field data of both NbAs and NbP, we assume this to be the likely cause of their LMR for $B||a$.

\section{Conclusion}

The aim of this study was the achievement of a homogeneous current distribution in the measurement of the longitudinal magnetoresistance of the TaAs-family of Weyl metals. We achieved this in all materials but TaAs and found especially the $c$-axis transport to exhibit a strictly positive LMR. While the $a$-axis transport in TaP strongly resembled the $c$-axis counterpart, both niobium compounds showed a negative, or even non-monotonic, behaviour.

Our second intention was to interpret the data without the strong expectation to find the chiral anomaly. Indeed, we found a number of possible effects which may exhibit a negative LMR and which have not been ruled out so far in the literature. 

We found the $c$-axis transport in all compounds to be dominated by an orbital magnetoresistance. This is in full agreement with calculations of the expected size of the orbital LMR based on the Fermi surface corrugation and the fit results to the LMR.

In contrast, the $a$-axis transport in both niobium compounds is more likely recovered by a WAL model then a chiral anomaly. Given the qualitative similarity of the $a$-axis transport in TaAs and the excess of positive LMR found here, we believe it to be crucial to rule out the WAL scenario before any claims regarding the observation of a chiral anomaly can be made.



\section{Acknowledgements}

The authors would like to acknowledge the Max-Planck society for their support of the MPRG. EH and MN acknowledge support from the DFG through the project TRR80 "From electronic correlations to functionality".

\end{document}